# Thermal Conductivity of Industrial Nb$_3$Sn Wires Fabricated by Various Techniques.

M. Bonura, and C. Senatore

*Abstract*—We have developed a new experimental setup specifically designed for measuring thermal conductivity on technical superconductors in the range of temperatures 3–330 K in magnetic fields up to 21 T. Three Nb$_3$Sn wires, produced by the powder-in-tube technique, the bronze route and the internal tin restacked-rod process, respectively, have been investigated. We show that, due to the complexity of the architecture of these wires, direct measurement of thermal conductivity is required for a correct estimation of thermal stability in superconducting magnets.

## I. INTRODUCTION

THERMAL STABILITY is one of the most important issues when designing a superconducting magnet. In order to increase the magnetic field magnitude, $B$, as well as the bore diameter, higher operating currents are needed. The energy stored in a magnet scales with $B^2$ and with the magnet volume. In case of thermal disturbance this huge magnetic energy can be easily converted into heat. If the superconductor (SC) temperature rises locally above its critical value and the heat generation due to Joule effect overcomes the heat evacuated by the coolant, the normal zone will expand and eventually the whole magnet will revert to the normal state. This process is referred as a quench.

Among the industrial superconductors, Nb$_3$Sn wires are the most used in magnet technology. For this reason the study of their thermal stability is of basic interest [1,2]. The critical current density ($J_c$) in a SC decreases with increasing temperatures. If $T_0$ and $J_0$ are the operating temperature and current density of a superconducting device, respectively, dissipations arise at the temperature $T_g$ defined by the condition $J_c(T_g) = J_0$. However, in case of point disturbances, another condition must be fulfilled in order to get a quench: the dimension of the over-heated volume has to be wider than the so called minimum propagation zone (MPZ). The length of the MPZ in a superconducting wire can be estimated by the following formula:

$$l_{MPZ} \approx \sqrt{\frac{2k(T_c - T_0)}{J_c^2 \rho}} \qquad , \qquad (1)$$

where $k$ is the thermal conductivity of the wire and $\rho$ is the electrical resistivity [3]. If the initial perturbation develops in a region smaller than the MPZ, the local overheating will be reabsorbed by heat conduction along the wire. Once defined the operating conditions ($T_0$, $J_0$) it follows from (1) that thermal stability can be only improved by increasing the ratio $k/\rho$.

We have developed an experimental setup specifically designed for measuring the thermal conductivity of superconducting wires and tapes used in applications. The system operates in the range of temperatures 3–330 K with applied magnetic fields up to 21 T. In the present paper we report a study of the thermal properties of Nb$_3$Sn multifilamentary round wires prepared by different techniques, namely powder-in-tube (PIT) technique, bronze route (BR) and internal tin (IT) rod-restack process. Thermal conductivity has been investigated with this setup at zero magnetic field as well as at $\mu_0 H \approx 15$ T. Important information on the thermal properties of the wires has been obtained. In particular, we present a comparison between the experimental measurements of thermal conductivity and the theoretical prediction based on the residual resistivity ratio (RRR) measurement. The reliability of the RRR as indicator for the thermal properties of the composite Nb$_3$Sn wires is also discussed.

## II. EXPERIMENTAL SETUP AND SAMPLE

### A. The New Experimental Setup

The new experimental setup allows measuring thermal conductivity by means of the steady-state axial heat flow method. Heat power, $Q$, is supplied at one end of the sample and a gradient of temperatures is generated along the wire. From the Fourier-Biot law:

$$Q = -kA \frac{dT}{dx} \qquad , \qquad (2)$$

where $k$ is the thermal conductivity, $A$ is the cross-section of the sample, and $dT/dx$ is the temperature gradient. Experimentally, $dT$ is approximated by $\Delta T \equiv (T_2 - T_1)$, the finite temperature difference measured by means of two Cernox® bare-chip thermometers, and $dx$ is approximated by $\Delta x$, the distance between the thermometers. The thermal conductivity determined by this method is the average value in the temperature range $T_1$–$T_2$ (typically $\Delta T \approx 0.1$K for our setup).

Fig. 1 shows a sketch of the central part of the experimental probe. Particular attention has been devoted to the reduction of all possible error sources. The entire probe is inserted in a high-vacuum ($< 10^{-5}$ mbar) recipient with low-emissivity ($\approx 0.018$) walls in order to make the convective and the radiant losses negligible. The contact between the hot-terminal, where the heater is placed, and the rest of probe is limited to two 0.1 mm diameter nylon wires. All this allows us to evaluate that the fraction of heat $Q$ generated by the heater that is not transmitted through the sample is smaller than the 0.05%, even for samples with $k$ values as low as 1 Wm$^{-1}$K$^{-1}$. All the thermometers installed in the probe have been calibrated at zero field as well as in magnetic fields up to 21 T. In our setup it is possible to measure the thermal conductivity both for fields parallel and perpendicular to the sample axis. This allows studying anisotropy effects when needed.

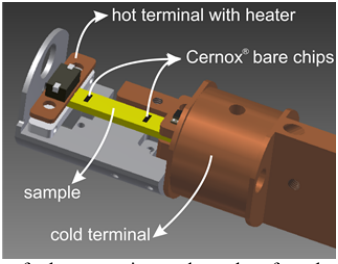

Fig. 1. Sketch of the experimental probe for thermal conductivity measurements.

In order to verify the performance of the experimental setup, measurements on copper samples of different purity have been performed. The copper temperature dependence of $k$ can be described by using only one parameter, namely the residual resistivity ratio, $RRR$ [4]. In particular, the experimental results at zero magnetic field can be predicted within ± 15% by the following expression:

$$k_{Cu} = (W_0 + W_i + W_{io})^{-1}, \qquad (3)$$

Where

$$W_0 = \frac{\beta}{T}$$

$$W_i = \frac{P_1 T^{P_2}}{\left(1 + P_1 P_3 T^{(P_2+P_4)} e^{-\left(\frac{P_5}{T}\right)^{P_6}}\right)}, \qquad (4)$$

$$W_{i0} = P_7 \frac{W_i W_0}{W_i + W_0}$$

with $\beta \approx 0.634/RRR$, $P_1 = 1.754 \times 10^{-8}$, $P_2 = 2.763$, $P_3 = 1102$, $P_4 = -0.165$, $P_5 = 70$, $P_6 = 1.756$, $P_7 \approx 0.235\ RRR^{0.1661}$ in SI units [4].

Fig. 2 shows the temperature dependence of thermal conductivity of a round copper wire measured in the range of temperatures from 5 K to 80 K. The continuous line is the best fit curve obtained by a least squares fitting of the experimental results using (3) and (4), leaving $RRR$ as free parameter. The best fit value obtained for $RRR$ is 24. This value is very compatible with the one deduced by resistivity measurements on a sample extracted from the same batch, whose $RRR$ is 26.

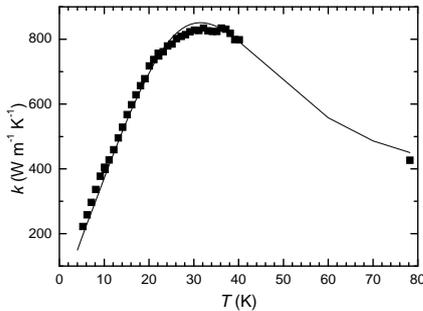

Fig. 2. Temperature dependence of thermal conductivity of a copper wire. Continuous line is the best fit curve obtained as detailed in the text.

### B. Samples

We have investigated three Nb$_3$Sn multifilamentary wires produced by different techniques, namely powder-in-tube method [5], bronze route [6] and internal tin rod restack process [7]. Samples have been labelled as PIT#1, BR#1 and IT#1, according to the fabrication technique used. Wire BR#1 has been produced at the Group of Applied Physics of the University of Geneva. PIT#1 and IT#1 are commercial wires produced by Bruker EAS and OST, respectively. Relevant

|  | PIT#1 | BR#1 | IT#1 |
|---|---|---|---|
| Diameter (mm) | 1 | 0.96 | 0.9 |
| Number of filaments | 192* | 14641 | 54* |
| $s_{Cu} = S_{Cu}/S_{tot}$ | 0.56 | 0.16 | 0.39 |
| Non-Cu $J_c$ [$T = 4.2K$, $H = 15T$] (A/mm$^2$) | 1374 | 432 | 699 |
| $RRR$ (from resistivity) | 330 | 270 | 180 |
| $RRR$ (from $k_{Cu}$ fit, Sec. III) | 238 | 487 | 65 |

\* Number of subelements

parameters of the samples are listed in Table I. Critical current density values have been measured at $\mu_0 H = 15$T using the 0.1 $\mu$V/cm criterion. $RRR$ values have been deduced from resistivity measurements performed at low as well as at room temperatures. Copper fractions have been provided by manufacturers. Transverse cross-sections of the investigated wires are shown in Fig. 3.

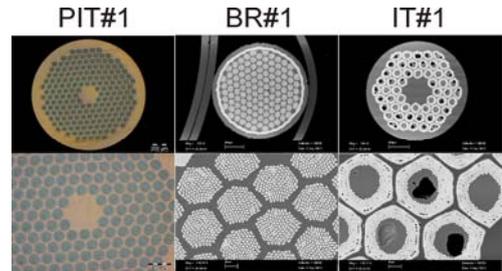

Fig. 3. Micrographies of the transvers cross sections of investigated wires.

### III. RESULTS AND DISCUSSION

Thermal conductivity of the three Nb$_3$Sn conductors has been investigated as a function of temperature, in the range 5–40 K, at zero magnetic field as well as at 15 T. Experimental results are shown in Fig. 4. All the $k(T)$ curves exhibit a maximum whose value and position depend on the specific sample investigated. In particular, at zero field the maximum of $k$ occurs at $T = 15$ K for sample PIT#1, at $T = 14$ K for sample BR#1 and at $T = 20$ K for sample IT#1. A clear reduction of $k$ is observed when applying a magnetic field of 15 T.

The heat conduction in composite superconducting wires is normally dominated by the matrix contribution. For polycrystalline Nb$_3$Sn, the thermal conductivity at 20 K is $k \approx 2$ Wm$^{-1}$K$^{-1}$ [8]. This value is more than two orders of magnitude smaller than what expected for copper with a *RRR* of 30 ($k \approx 860$ Wm$^{-1}$K$^{-1}$ at $T = 20$ K, $H = 0$). At temperatures smaller than $T_c$ the thermal conductivity of the superconductor is even smaller since Cooper pairs do not transmit heat [9].

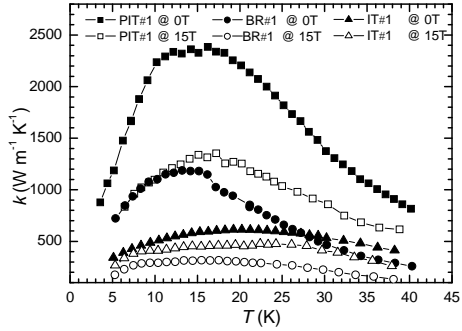

Fig. 4. Thermal conductivity as a function of temperature for the investigated wires. Full and open symbols indicate data acquired at $H= 0$ T and $\mu_0 H = 15$T, respectively.

In a metal, heat conduction can be considered as the sum of the electronic ($k_e$) and the lattice contributions ($k_l$), $k = k_e + k_l$. In pure metals, the electronic contribution is predominant being $k_l/k < 0.2$ [4,10]. The electronic contribution $k_e$ is in turn determined mainly by two mechanisms, a) the interaction of electrons with phonons, b) the interaction of electrons with defects. Concentrating on copper, usually used as thermal stabilizer in the wires, the temperature dependence of thermal conductivity is described by (3) and (4). At low temperature, $k$ depends strongly on *RRR*. In particular, large *RRR* implies high $k$ values. It follows that copper purity is crucial for the thermal stability of the wire. Also the position of the maximum in the $k_{Cu}(T)$ curve depends on the metal purity [4], shifting towards lower temperatures on increasing the metal purity. At temperatures above 100 K, $k$ is almost independent of *RRR*. In this case, the largest part of the thermal resistance comes from electron-phonon interactions, making the amount of defects less important.

In case of multicomponent wires, the measured longitudinal thermal conductivity, $k_{meas}$ is the sum of the thermal conductivity of each component, $k_i$, weighted for the corresponding surface fraction:

$$k_{meas} = \sum_i k_i \frac{S_i}{S_{tot}} = \sum_i k_i s_i \ . \tag{5}$$

$S_i$ and $S_{tot}$ are the surface of the wire cross section occupied by the $i^{th}$ component and the total cross section area of the wire, respectively. Typical $k$ values of Nb$_3$Sn wires materials [8] imply that the main contribution to heat transport is due to the stabilizing copper. As a consequence, we have deduced the temperature dependence of the thermal conductivity of the copper present in the samples, $k_{Cu}$, by supposing that $k_{meas} \approx k_{Cu} s_{Cu}$. In this case, one is also assuming that local variations of the copper purity can be neglected, i.e. there is not a distribution of *RRR* values in the wire.

Fig. 5 shows the temperature dependence of thermal conductivity of the copper present in the wires, $k_{Cu}$. Points are values obtained by dividing the experimental data $k_{meas}$ by $s_{Cu}$ values given by manufacturers and reported in Table I. Dashed lines are the theoretical expectations for $k_{Cu}$ calculated by using (3) and (4) and the *RRR* obtained by resistivity measurements, listed in Table I. Continuous lines are the best-fit curves obtained using (3) and (4) with *RRR* as free parameter. The obtained best-fit values for *RRR* are: $RRR = 238$, $RRR = 487$ and $RRR = 65$ for samples PIT#1, BR#1 and IT#1, respectively. *RRR* values obtained from the fit analysis have also been listed in Table I. This for an easier comparison with values obtained from resistivity measurement.

It is interesting to note the large discrepancy between the experimental thermal conductivity values of the copper composing the wires and the theoretical expectations calculated by using the measured *RRR*. A much better agreement is obtained by fitting the experimental data treating *RRR* as free parameter. However, fit results are less satisfactory when compared to what obtained in case of the pure copper wire (see Fig. 2). This is most likely related to the approximations made when deducing $k_{Cu}$ from the measured thermal conductivity of the whole wire. Experimental data and best fit curve are particularly discrepant for the IT wire. In our opinion this could be due to the local inhomogeneity of copper purity. In fact, especially in case of the internal-tin technique, outer copper can be much purer than copper inside the filaments, which can encounter contamination due to tin diffusion from subelements. Radial distribution of *RRR* can play a role also for the PIT wire. We believe that a satisfactory fit for all the investigated wires could be achieved by considering a proper *RRR* distribution in the wire as well as a very accurate determination of the Cu/non-Cu ratio and possible second order contributions to the overall $k_{meas}$ coming from other components. In particular, the contribution of Nb to the thermal conductivity of the wire could be in general not negligible. In fact, thermal properties of Nb depend strongly on its purity. For example, thermal conductivity of pure niobium is very high, resulting $k \approx 400$ Wm$^{-1}$K$^{-1}$ at $T = 10$ K for $RRR = 300$ [8]. However, a full comprehensive analysis leading to a more accurate fit of the $k_{Cu}(T)$ curves goes beyond the aim of the present paper.

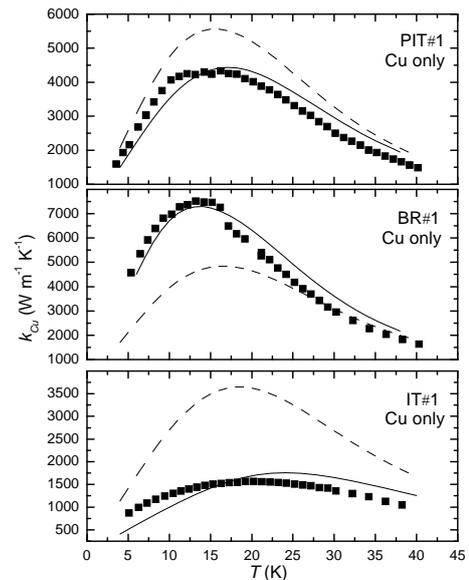

Fig. 5. Temperature dependence of thermal conductivity of copper, $k_{Cu}$, in Nb$_3$Sn wires. Points indicate values deduced from the experimental results. Dashed lines are the theoretical curves calculated using measured *RRR*. Solid lines are the best fit curves obtained as described in the text.

Thermal and electrical conductivity of pure metals are closely connected parameters, being associated with the same conduction source and limiting mechanisms [10]. In case of single component wires, the electrical conductivity is often used to estimate $k$ since it is much easier to measure. On the contrary much more attention should be paid in case of multicomponent wires. In absence of direct thermal measurements, thermal properties of the wire at low temperatures can be estimated from the $RRR$ of the stabilizing metal, typically copper. However, standard resistivity measurements provide the $RRR$ of the entire composite.

Limits in deducing thermal conductivity from the $RRR$ of the wire are clear when analysing the experimental results obtained in sample BR#1. In this case we have $RRR_{wire} = 270$ from the electric resistivity measurements and $RRR_{Cu} = 487$ from fitting the thermal conductivity data. This difference is very important when designing magnets since $k_{Cu}(4.2 \text{ K}) = 1780$ Wm$^{-1}$K$^{-1}$ for copper with $RRR = 270$ whereas $k_{Cu}(4.2 \text{ K}) = 3210$ Wm$^{-1}$K$^{-1}$ with $RRR = 487$, i.e. 80% higher.

Our analysis suggests that the estimation of thermal properties from the $RRR$ of the wire can be misleading. $RRR$ values of the entire wire extracted from resistivity measurements may differ considerably from the $RRR$ values deduced for copper by fitting the $k_{Cu}(T)$ curves. Both sets of values are listed in Table I. The obtained result can be easily interpreted in case of the BR wire. Considering the thermal and electrical conductivities of bronze and copper [4,8], one can deduce that the bronze does not play a role at low temperatures neither for heat nor for electric transport. The situation is different at room temperature when the fraction of electric current that passes through the bronze matrix influences the overall resistivity. This can lead to measured $RRR$ values of the strand smaller than that for copper only. Consequently, copper purity and expected $k$ values at low temperatures can be underestimated if proper considerations are not done. In case of the PIT#1 and IT#1 samples, the measured $RRR$ values would lead to an over-estimation of the thermal conductivity. This result cannot be easily understood as in case of the BR#1 sample, due to the higher complexity of the conductor architecture. This further confirms that a simple correspondence between electrical and thermal properties is not always possible.

Our analysis has allowed us to conclude that the copper purity of the BR#1 sample is higher than that of the PIT#1 sample, in spite of $RRR$ values obtained by resistivity measurements. This conclusion coming from the fit of the experimental data obtained at $H = 0$ is further confirmed by analysing the experimental results obtained at 15 T. A reduction of copper thermal conductivity is expected by applying a magnetic field. The variation is more pronounced at higher $RRR$ values [11]. The effects of the magnetic field on $k$ for the examined superconducting wires are shown Fig. 4. If we define $R \equiv k_{max}(0\text{T})/k_{max}(15\text{T})$ we obtain $R = 1.8$ for PIT#1, $R = 3.8$ for BR#1 and $R = 1.3$ for IT#1. $R$ values indicate that the reduction is much more pronounced for the wire produced by the bronze-route technique, confirming the higher copper purity in BR#1. The variation of thermal conductivity on applying a magnetic field is thus a useful cross check for estimating the copper purity level in composite wires.

From the analysis of the overall thermal conductivity data reported in Fig. 4, one can observe that the $k(T)$ curve of the PIT#1 wire is clearly higher than those of other samples. This occurs even if the purity of the composing copper is lower than that of the BR#1 wire. This result is a consequence of the larger Cu/non-Cu fraction for PIT#1 compared to the other samples. The design of the PIT#1 wire gives also the advantage of limiting the reduction of $k$ on applying magnetic fields. Even for $\mu_0 H = 15$T, the thermal conductivity of the PIT#1 is higher than those measured in BR#1 and IT#1 at zero field.

As discussed in the introduction, thermal conductivity is an essential parameter for modelling quench processes in superconducting magnets. In order to give an idea of how differences in thermal conductivity can affect the performance of a superconducting magnet, we have estimated the minimum-propagation-zone length, $l_{MPZ,}$ at $T_0 = 4.2$ K for all the investigated wires. From (1), considering the measured values for thermal conductivity, electrical resistivity and critical current density at $\mu_0 H = 15$ T, one obtains: $l_{MPZ} \approx 3$ mm for PIT#1 wires and $l_{MPZ} \approx 2$ mm for BR#1 and IT#1 wires. The slightly wider MPZ for PIT#1, in spite of the higher sustainable current density, is a consequence of the higher thermal conductivity of the PIT#1 wire in comparison to the others.

## IV. Conclusion

A new experimental set-up specifically designed for studying thermal properties of superconducting wires and tapes has been developed. The temperature and field dependence of thermal conductivity in Nb$_3$Sn wires produced by different techniques has been investigated. The analysis of the experimental results has allowed us to deduce important information about the purity of copper present in the wire after the reaction. In particular, we have evaluated the error made when estimating thermal conductivity at low temperature starting from the overall $RRR$ of the wire. The minimum-propagation-zone length, important parameter for preventing quench processes in superconducting magnets, has been estimated for the 3 different wires.

## Acknowledgment

We acknowledge Alexandre Ferreira for his precious support when designing the experimental probe, Florin Buta for very useful discussions and Damien Zurmuehle for technical assistance.